\documentclass[showkeys,showpacs]{revtex4}
\usepackage{amsmath}
\usepackage{amssymb}
\usepackage{graphicx,color}

\begin{document}

\title{On the self-similar motion of a gravitating Chaplygin fluid}

\author{Vladimir Dzhunushaliev,$^{1,2,3}$
\footnote{
Email: vdzhunus@krsu.edu.kg}
Vladimir Folomeev$^{3}$
\footnote{Email: vfolomeev@mail.ru}
and
Ratbay Myrzakulov$^{4}$
\footnote{Email: cnlpmyra1954@yahoo.com, cnlpmyra@mail.ru}
}
\affiliation{
$^1$
Institut f\"ur Physik, Universit\"at Oldenburg, Postfach 2503
D-26111 Oldenburg, Germany \\
$^2$Department of Physics and Microelectronic
Engineering, Kyrgyz-Russian Slavic University, Bishkek, Kievskaya Str.
44, 720021, Kyrgyz Republic \\ 
$^3$Institute of Physicotechnical Problems and Material Science of the NAS
of the
Kyrgyz Republic, 265 a, Chui Street, Bishkek, 720071,  Kyrgyz Republic \\
$^4$ Department of General and Theoretical Physics,
Eurasian National University, Astana, 010008, Kazakhstan
}

\begin{abstract}
A self-similar motion of the generalized Chaplygin gas in its own gravitational field
 is considered. The problem is  studied numerically and analytically (in the limiting cases).
 It is shown that the model under consideration admits only expanding solutions.
 As the astrophysical application of the model, a description of rotating curves of
 spiral galaxies is suggested by using the analytical solutions obtained in the paper.
\end{abstract}

\pacs{97.10.-q, 95.35.+d, 98.35.Gi}
\keywords{Similarity, Chaplygin gas, dark fluid}

\maketitle

\section{Introduction}

The problem of motion of a fluid in its own gravitational field (collapse)
 refers to key problems of astrophysics.
Until the end of the 1990's,  studies of models of compact objects and a collapse  process have been basically concentrated on
a consideration  of models in which a fluid possesses a positive (or zero) pressure.
However, the situation has changed after the discovery of the accelerated expansion of the present universe
 \cite{Perlmutter:1998np,Riess:1998cb}. It is now believed that such acceleration is caused by the presence in the universe
of some antigravitating matter with the negative pressure
$p<-1/3\, \varepsilon$, where $\varepsilon$ is the energy density
 (for a review, see e.g. \cite{Sahni:2004ai,Sahni:2006pa}). It is assumed that such matter, which is called dark energy, is
distributed uniformly in space and provides more than 70\% of a total mass of the universe.
However, a possibility is not excluded that dark energy could affect the formation of the structure of the
universe on small scales of the order of galaxies, and may be even smaller (for a more detailed discussion of this question,
see Ref. \cite{Mota:2004pa}).

A more intriguing possibility is a description of dark energy and dark matter within the framework of some unified theory by
using an effective hydrodynamical description of a dark fluid. One of variants of such a theory is a model of the Chaplygin gas
\cite{Kamenshchik:2001cp}.
It is assumed in this model that the dark fluid is being described by the polytropic equation of state
 $p_{ch}\sim -\rho_{ch}^{\gamma}$, where
$p$ and $\rho$ are the pressure and the energy density, respectively, and
 $\gamma$ is a negative constant. In a cosmological aspect, such equation of state gives the following scenario of the evolution
 of the universe: in the early stages of the expansion of the universe
 $\rho_{ch}\sim a^{-3}$ (here $a$ is the scale factor) that corresponds to the epoch of domination of
 non-relativistic (dustlike) dark matter. On the other hand, at the present time and later on such
fluid acts as a cosmological constant, $\rho_{ch}\sim const$. This attractive feature allows to use such model
for a description of the evolution of the universe from the epoch of formation of the large scale structure  until
the present time.

Possessing the properties of dark matter, the Chaplygin gas can form compact astrophysical objects
such as
stars and galaxies. Such objects had been already considered earlier. In particular, in Ref.~\cite{Bertolami:2005pz}
 the model of the dark energy star supported by the Chaplygin gas was considered within the framework of Newtonian gravity.
Choosing various values of a central density of such configuration, sizes and masses of dark stars in the early universe were
estimated. In general relativity, such stars were considered in detail in Refs.~\cite{Gorini:2008zj,Gorini:2009em}.

In this paper we consider the motion of the Chaplygin gas in its own gravitational field within the framework of Newtonian gravity.
Such model can be used for a description of evolution of the dark fluid inside galaxies. Since the sizes of
such objects are significantly smaller than cosmological scales then one can use the Newtonian approximation. In the paper
\cite{Bertolami:2005pz}  the parameters of the Chaplygin dark star were estimated, taking into account the cosmological
evolution of the Chaplygin gas.
Here we will estimate  parameters of a motion of the Chaplygin gas
creating a compact astrophysical object.
To do this, we will consider a self-similar motion of the gas with the above equation of state. Such problems
for self-gravitating gaseous spheres have been already considered by a number of authors
(see e.g. earlier works \cite{Shu:1977uc,Bouquet:1985,Suto:1988}, and also more recent papers \cite{Lou:2006jn,Wang:2008rs}
and references therein).
In these works the main attention was called to the models of collapsing stars with the parameter
 $\gamma>0$. It will be shown below that in the case of
 $\gamma<0$ (the Chaplygin gas) self-similar solutions considered here describe only expanding configurations.

\section{Equations and solutions}

\subsection{Similarity equations}
We will consider the spherically symmetric motion of the Chaplygin gas in Newtonian gravity.
The corresponding hydrodynamical equations can be written in the form  \cite{Suto:1988}
\begin{eqnarray}
\label{hydro-1}
\frac{\partial M}{\partial t}+4\pi r^2 \rho u&=&0,
 \\
\label{hydro-2}
\frac{\partial M}{\partial t}&=&4\pi r^2 \rho,
 \\
\label{hydro-3}
\frac{\partial u}{\partial t}+u\frac{\partial u}{\partial r}&=&-\frac{1}{\rho}\frac{\partial p}{\partial r}-\frac{G M}{r^2},
\end{eqnarray}
where $u$ is the radial velocity of the Chaplygin gas,
$M$ is the total mass of matter contained in the interval $(0,r)$, $\rho$ and
$p$ are the density and the pressure of matter, respectively.
In the case under consideration, the pressure $p$ is small compared with the energy density $\rho$.
Equations  \eqref{hydro-1}-\eqref{hydro-3} are invariant under the time reversal operation \cite{Shu:1977uc}:
$$
t\to -t, \quad u\to -u, \quad \rho \to \rho, \quad p\to p, \quad M\to M.
$$
Thus one can consider only the range of $0<t<\infty$. Let us choose the
polytropic equation of state of the generalized Chaplygin gas in the form
\begin{equation}
\label{eqs_chapl}
p=-K \rho^\gamma,
\end{equation}
where $K$ is some positive arbitrary constant and $-1\leq \gamma <0$. The case $\gamma=-1$ corresponds to the
usual Chaplygin gas \cite{Kamenshchik:2001cp}.

In this paper, we will seek similarity solutions of the system of equations
 \eqref{hydro-1}-\eqref{hydro-3} with the equation of state
\eqref{eqs_chapl}. For this purpose, we introduce the following dimensionless similarity variables
(for details, see
 \cite{Suto:1988}):
\begin{align}
\label{self_variab}
\begin{split}
x&=\frac{r}{\sqrt{k}t^n},\quad u(r,t)=\sqrt{k}t^{n-1}v(x), \quad \rho(r,t)=\frac{\alpha(x)}{4\pi G t^2}, \\
p(r,t)&=-\frac{k}{4\pi G}\,t^{2n-4}[\alpha(x)]^\gamma, \quad M(r,t)=\frac{k^{3/2} t^{3n-2}}{(3n-2)G}\,m(x),
\end{split}
\end{align}
where $k$ is some dimension constant. Using these similarity variables, Eqs.
\eqref{hydro-1}-\eqref{hydro-3} can be rewritten in the form
\begin{eqnarray}
\label{hydro-1_self}
m(x)&=&\alpha x^2(n x-v),
 \\
\label{hydro-2_self}
(n x-v)\frac{d\alpha}{dx}-\alpha\frac{dv}{dx}&=&-2\frac{x-v}{x}\,\alpha,
 \\
\label{hydro-3_self}
-\gamma\alpha^{\gamma-2}\frac{d\alpha}{dx}-(n x-v)\frac{dv}{dx}&=&-(n-1)\,v-\frac{n x-v}{3n-2}\,\alpha.
\end{eqnarray}
The similarity variables \eqref{self_variab} in general suppose that
\begin{equation}
\label{par_K}
K=k(4\pi G)^{\gamma-1}t^{2(n+\gamma-2)}.
\end{equation}
If we restrict ourselves to the case of constant
 $K$ (as it is usually assumed at a consideration of the dark Chaplygin fluid),
the parameter
$n$ must be restricted by the additional condition
\begin{equation}
\label{par_n}
n=2-\gamma,
\end{equation}
or, taking into account \eqref{eqs_chapl}, $2<n\leq 3$.
For convenience of performing the mathematical analysis, let us rewrite Eqs.
\eqref{hydro-2_self} and \eqref{hydro-3_self} in the following form
\begin{eqnarray}
\label{hydro-2_self_n}
\frac{d\alpha}{dx}&=&\frac{\alpha}{(n x-v)^2+\gamma\alpha^{\gamma-1}}
\left[(n-1)v+\frac{n x-v}{3n-2}\,\alpha-2\frac{(x-v)(n x-v)}{x}\right],
 \\
\label{hydro-3_self_n}
\frac{dv}{dx}&=&\frac{1}{(n x-v)^2+\gamma\alpha^{\gamma-1}}
\left[(n-1)(n x-v)v+\frac{(n x-v)^2}{3n-2}\,\alpha+2\,\gamma\,\frac{x-v}{x}\,\alpha^{\gamma-1}\right].
\end{eqnarray}

\subsection{Limiting behavior}
In general, the above equations correspond to a non-autonomous system which should be investigated numerically.
However, let us estimate first the behavior of the system
 \eqref{hydro-2_self_n}-\eqref{hydro-3_self_n} near the origin of coordinates. We will follow Ref. \cite{Suto:1988}:
we expand $\alpha$ and $v$ in a Taylor series in the neighborhood of $x=0$:
\begin{equation}
\label{Taylor}
\alpha(x)=\alpha_*+\alpha_1 x +\alpha_2 x^2+\cdots, \quad
v(x)=v_0+v_1 x+v_2 x^2+\cdots.
\end{equation}
Substituting these expressions in Eqs. \eqref{hydro-2_self_n} and \eqref{hydro-3_self_n}, one can obtain the
following limiting solutions for $x\ll 1$ parameterized by $\alpha_*$:
\begin{eqnarray}
\label{bound_zero_1}
v(x)&=&\frac{2}{3}x+\frac{\alpha_*^{1-\gamma}}{15\gamma}\left(\alpha_*-\frac{2}{3}\right)\left(n-\frac{2}{3}\right)x^3+\cdots,
 \\
\label{bound_zero_2}
\alpha(x)&=&\alpha_*+\frac{\alpha_*^{2-\gamma}}{6\gamma}\left(\alpha_*-\frac{2}{3}\right)x^2+\cdots.
\end{eqnarray}
Note that Eqs. \eqref{hydro-2_self} and \eqref{hydro-3_self} have a particular solution with the constant density
 $\alpha$
\begin{equation}
\label{homogen}
v=\frac{2}{3}x, \quad \alpha=\frac{2}{3}, \quad m=\frac{2}{3}\left(n-\frac{2}{3}\right)x^3
\end{equation}
corresponding to the homogeneous distribution of matter and
describing  expansion in Newtonian cosmology.

\begin{figure}[ht]
\centering
  \includegraphics[height=10cm]{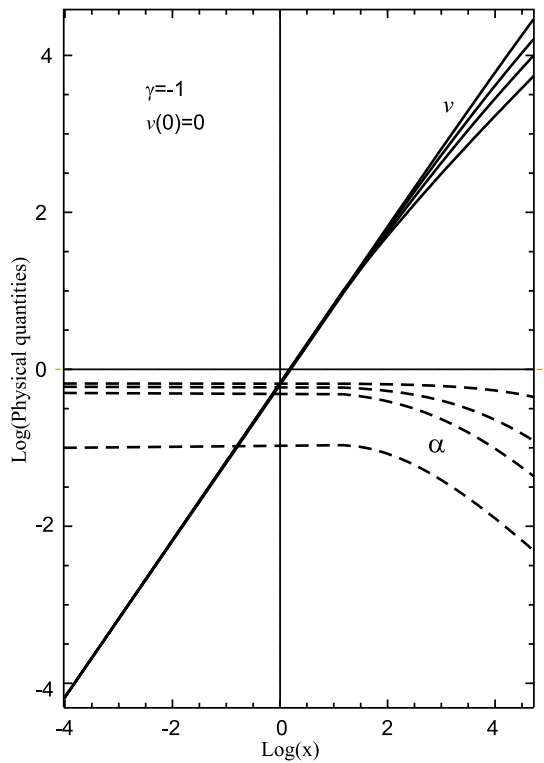}
\caption{The  graphs of  velocities
$v(x)$ (the solid lines)  and densities
$\alpha(x)$ (the dashed lines) at  different initial values:
$\alpha_*=0.66, 0.6, 0.5, 0.1$, from top to bottom, for both sets of the curves.
}
    \label{veloc}
\end{figure}

Using \eqref{bound_zero_1},\eqref{bound_zero_2} as initial conditions, let us solve  Eqs.
\eqref{hydro-2_self_n}, \eqref{hydro-3_self_n} numerically. The results are presented in Fig.~\ref{veloc}.

The numerical analysis indicates two regions of the solutions. In the first region
 $\alpha_*<2/3$. In this case there exists
 the critical point in which the numerators and denominators of
Eqs.
\eqref{hydro-2_self_n}, \eqref{hydro-3_self_n} vanish simultaneously.
The position of this point can be defined from the conditions:
\begin{eqnarray}
\label{crit_point_1}
&&(n x-v)=\sqrt{-\gamma}\alpha^{(\gamma-1)/2},
 \\
\label{crit_point_2}
&&(n-1)v+\frac{n x-v}{3n-2}\alpha-2\frac{(x-v)(n x-v)}{x}=0.
\end{eqnarray}
An asymptotic form of solutions of Eqs.
 \eqref{hydro-2_self_n}, \eqref{hydro-3_self_n} at $x\gg 1$ is defined from the form of the numerical solutions.
One can see from the numerical calculations that
$v$ changes more slowly than
 $x$, and the function
$\alpha^{\gamma-1}$ changes more slowly than
 $x^2$. It allows to find the following asymptotic solutions: from Eq.
\eqref{hydro-2_self_n} we have
\begin{equation}
\label{asymp_alpha}
\frac{d\alpha}{dx}=-\frac{2\alpha}{n x} \quad \Rightarrow \quad \alpha=A x^{-2/n}.
\end{equation}
Using this solution, Eq.
 \eqref{hydro-3_self_n} can be rewritten as
$$
\frac{dv}{dx}=\frac{(n-1)v}{n x}+\frac{A}{3n-2}x^{-2/n}+\frac{2\gamma}{n^2}A^{\gamma-1}x^{-2(\gamma-1+n)/n}
$$
with solution
$$
v=B x^{(n-1)/n}-\left\{\frac{n A}{3n-2}+\frac{2\gamma A^{\gamma-1}}{\left[2(n+\gamma)-3\right]n}\right\}x^{-\gamma/(2-\gamma)}.
$$
Here $A,B$ are integration constants. Taking into account the relation
 $n=2-\gamma$, one can see from the last equation that a dominating term is the term with
 $B$ (if, of course, one omits the case
$B=0$). Then we have
\begin{equation}
\label{asymp_v}
v=B x^{(n-1)/n}.
\end{equation}
Taking into account the restrictions on values of
 $-1\leq \gamma <0, \,\,2<n \leq 3$, one can see from Eqs.
\eqref{asymp_alpha} and \eqref{asymp_v} that the density of the Chaplygin gas is a slowly decreasing function at
 $x\gg 1$: $\alpha\sim x^{-2/3}$ at $\gamma=-1$ and
$\alpha \sim x^{-1}$ at $\gamma \to -0$. At the same time, the velocity
$v$ is an increasing function: $v\sim x^{2/3}$ at $\gamma=-1$ and
$v \sim x^{1/2}$ at $\gamma \to -0$.

In the second region of the solutions, when
$\alpha_*>2/3$, there also exists the critical point which is defined by  equations
 \eqref{crit_point_1} and
\eqref{crit_point_2}. However, in this case, somewhere on the solution plane of the system of equations
 \eqref{hydro-2_self_n}-\eqref{hydro-3_self_n}, there also exists a point where the denominator of Eq.
\eqref{hydro-2_self_n} goes to zero but the numerator is not equal to zero.
I.e. the solution becomes singular, and the function
 $\alpha$, as opposed to  the region
 $\alpha_*<2/3$, is not a decreasing function but an increasing one that is also physically unacceptable.

Because of the above, let us use  the regular similarity solutions
\eqref{bound_zero_1}, \eqref{bound_zero_2},
\eqref{asymp_alpha} and \eqref{asymp_v} in the region
 $\alpha_*<2/3$ for purposes of estimation of the physical quantities
 \eqref{self_variab} rewritten via the variables
 $(r,t)$. Using the definition for
 $x$ from \eqref{self_variab}, one can see that the limiting case $x\gg 1$ corresponds to the initial instant
$t\to0$  (with the arbitrary finite values of
$r$). Then, using the expressions
 \eqref{asymp_alpha}, \eqref{asymp_v}, we have from \eqref{self_variab}
\begin{equation}
\label{init_distr}
u(r,t)=B k^{1/2n}r^{(n-1)/n}, \quad \rho(r,t)=\frac{A k^{1/n}}{4\pi G}r^{-2/n}.
\end{equation}
I.e. at the initial instant there is the static inhomogeneous distribution of the Chaplygin gas. On the other hand,
the limiting case $x\ll 1$ corresponds to
 $t\to \infty$ at some finite $r$. The corresponding velocity and density of matter from
 \eqref{self_variab} will then be
\begin{equation}
\label{asymp_distr}
u(r,t)=\frac{2}{3}\frac{r}{t}, \quad \rho(r,t)=\frac{\alpha_*}{4\pi G t^2}.
\end{equation}
I.e. asymptotically, as
 $t\to \infty$, the model describes the homogeneous distribution of matter which does not depend on the equation of state
(i.e. on
$n$). If one follows the motion of some layer of matter then after transition to the Lagrangian variables, equation \eqref{asymp_distr}
for the velocity
 $u$  can be easily integrated giving the following expression
$$
r=r_0\left(\frac{t}{t_0}\right)^{2/3},
$$
where $r_0$, $t_0$ are integration constants describing characteristic
parameters of the configuration. In cosmology language, this equation
corresponds to the motion of dust matter. I.e. within the framework
of the model under consideration, the Chaplygin gas, creating the compact
astrophysical object,  describes asymptotically, as $t\to \infty$, a cloud of the
dark fluid with the homogeneous distribution of matter with the dustlike
equation of state. In this case the initially inhomogeneous distribution of
matter \eqref{init_distr} smooths asymptotically as $t\to 0$.

\section{Chaplygin gas in galaxies}
Since the solutions
\eqref{init_distr} are self-similar, i.e. they are similar to each other at
different  times,  then one can use these solutions for a description of the
present distribution of the dark fluid in the universe.
Since the Chaplygin gas may mimic dark matter, we want to apply the solution
obtained above as a dynamical model of dark matter inside a spiral galaxy. We
assume that dark matter inside a galaxy is a dynamical object evolving in
 time. We will apply the asymptotic solution \eqref{init_distr} for a
description of the dark matter distribution in a galaxy for $x \gg 1$
(that corresponds to $t={\rm const.}$, $r \gg \sqrt{k} t^n$). In our model a dark matter halo evolves in
 time, and we try to describe such evolution at some instant.
As an example, let us consider a possibility in the use of the solutions
\eqref{init_distr} for modeling dark matter in spiral galaxies. For such
galaxies there is
 the so-called  Universal Rotation Curve defining the velocities of the motion
of stars in a galaxy
  at any radius. The form of such curve can be represented by the following expression
 \cite{Persic:1995ru}
\begin{equation}
\label{URC}
V_{URC}(\tilde{x}) ~=~ V(R_{opt})~ \biggl[ \biggl(0.72+0.44\, {\rm log}
{L \over L_*}\biggr) ~{1.97~\tilde{x}^{1.22}\over{(\tilde{x}^2+0.78^2)^{1.43}}}~+
~
1.6\, e^{-0.4(L/L_*)}  { \tilde{x}^2 \over \tilde{x}^2+1.5^2 \,\left({L/ L_*}\right)^{0.4} } \biggr]^{1/2}
~~~~ {\rm km~s^{-1}},
\end{equation}
where $\tilde{x}=r/R_{opt}$,  $R_{opt}$ is the radius encompassing 83\% of the total integrated light,
$L$ is  the galaxy luminosity, and $\log{L_*/L_\odot}=10.4$.
As was pointed out in Ref.
\cite{Persic:1995ru},  Eq.~\eqref{URC} predicts rotation velocities at any  radius with a typical accuracy of 4\%.
The first term in this formula gives
 the contribution from a stellar disc, and the second one --  the contribution from a dark halo.
For our purposes, let us use the second term. Then we have from
 \eqref{URC}  the following  distribution of velocities which is defined by the dark halo
\begin{equation}
\label{URC_DM}
V_{DM}^{obs}(r/R_{opt}) ~=~ V(R_{opt})~ \sqrt{
 1.6\, e^{-0.4(L/L_*)} { \tilde{x}^2 \over \tilde{x}^2+2.25 \,\left({L/ L_*}\right)^{0.4} } }
~~~~ {\rm km~s^{-1}}.
\end{equation}

\begin{figure}[t]
\centering
  \includegraphics[height=5cm]{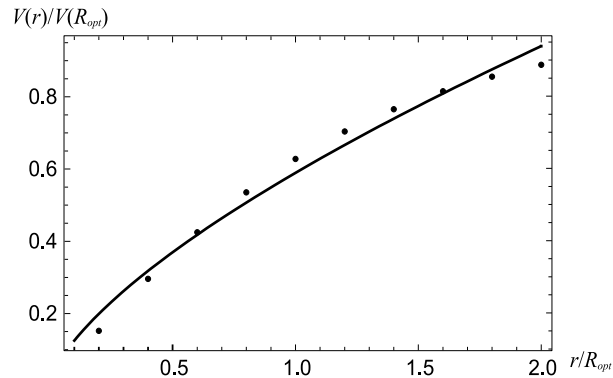}
\caption{Comparison of observational data
\eqref{URC_DM} (the dots) with the  curve
\eqref{V_Chapl_DM} for the Chaplygin gas
(the solid line) with the parameters $\tilde{D}\approx 0.59$ and $m\approx 0.67$.
}
    \label{fitting}
\end{figure}

Let us now estimate velocities of test particles moving in the field of gravity of the dark halo created by the Chaplygin gas.
The corresponding profile of the velocity distribution will then be given by the formula
$$
V_{DM}^{Chapl}(r)=\sqrt{\frac{G M(r)}{r}}.
$$
The expression for the mass distribution of the dark fluid follows from Eq.
 \eqref{hydro-1_self}. Taking into account Eqs.
 \eqref{self_variab} and \eqref{init_distr}, the expression for the mass profile at
 $x\gg 1$ will be
$$
M(r)=\frac{A n k^{1/n}}{(3n-2)G}\,r^{(3n-2)/n},
$$
and the corresponding  velocity distribution is
$$
V_{DM}^{Chapl}(r)=D \,r^{(n-1)/n}, \quad D=\sqrt{\frac{A n k^{1/n}}{3n-2}}.
$$
For convenience of comparison of this expression with  formula
 \eqref{URC_DM}, let us rewrite it via the variables
$\tilde{x}$ and $R_{opt}$ as
\begin{equation}
\label{V_Chapl_DM}
V_{DM}^{Chapl}(\tilde{x})=V(R_{opt}) \tilde{D} \tilde{x}^m, \quad \text{where} \quad \tilde{D}=\frac{D R_{opt}^{(n-1)/n}}{V(R_{opt})},
\quad m=\frac{n-1}{n}.
\end{equation}
Further, choosing the parameters
$\tilde{D}$ and $m$, one can compare Eq. \eqref{V_Chapl_DM} with  formula
\eqref{URC_DM}. As an example, let us take
$L/L_*=0.75$. The corresponding parameters will then be:
 $\tilde{D}\approx 0.59$ and $m\approx 0.67$
(see Fig.~\ref{fitting}). Since the parameter $\tilde{D}$ contains the integration constant
 $A$ then its value can be chosen arbitrarily. The value
 $m\approx 0.67$ corresponds to $n\approx 3$, i.e. $\gamma\approx -1$
(see eq. \eqref{par_n}) that corresponds to the usual Chaplygin gas.

Summarizing the obtained results, we have considered the self-similar motion of the Chaplygin gas with the equation
of state  \eqref{eqs_chapl}.  For values of the parameters used in creating the cosmological models with the
Chaplygin gas ($-1\leq \gamma <0$), it was shown that only expanding solutions do exist. This is caused by the nature of the Chaplygin gas
having a negative pressure. Within the framework of the model under consideration, the numerical analysis was performed, and
also the analytical estimations of the behavior of the solutions were made for the limiting cases when the similarity variable $x\gg 1$ and $x\ll 1$
(see Eqs. \eqref{init_distr} and \eqref{asymp_distr}, respectively).
The case of
 $x\gg 1$, corresponding to the static distribution of the Chaplygin gas, has been applied for the
evaluation of the velocity profile  of stars in galaxies.
 Comparing the observational curve for a dark halo
 \eqref{URC_DM} and the curve for the Chaplygin gas
 \eqref{V_Chapl_DM}, it was shown that
 they are in good agreement with each other
 at the corresponding
 choice of the fitting parameters
$\tilde{D}$ and $m$ (see Fig.~\ref{fitting}).


\begin{thebibliography}{99}


\bibitem{Perlmutter:1998np}
  S.~Perlmutter {\it et al.}  [Supernova Cosmology Project Collaboration],
  Astrophys.\ J.\  {\bf 517}, 565 (1999)
  [arXiv:astro-ph/9812133].
\bibitem{Riess:1998cb}
  A.~G.~Riess {\it et al.}  [Supernova Search Team Collaboration],
  Astron.\ J.\  {\bf 116}, 1009 (1998)
  [arXiv:astro-ph/9805201].
\bibitem{Sahni:2004ai}
  V.~Sahni,
  Lect.\ Notes Phys.\  {\bf 653}, 141 (2004)
  [arXiv:astro-ph/0403324].
\bibitem{Sahni:2006pa}
  V.~Sahni and A.~Starobinsky,
  Int.\ J.\ Mod.\ Phys.\  D {\bf 15}, 2105 (2006)
  [arXiv:astro-ph/0610026].

\bibitem{Mota:2004pa}
  D.~F.~Mota and C.~van de Bruck,
  Astron.\ Astrophys.\  {\bf 421}, 71 (2004)
  [arXiv:astro-ph/0401504].
\bibitem{Kamenshchik:2001cp}
  A.~Y.~Kamenshchik, U.~Moschella and V.~Pasquier,
  Phys.\ Lett.\  B {\bf 511}, 265 (2001)
  [arXiv:gr-qc/0103004].

\bibitem{Bertolami:2005pz}
  O.~Bertolami and J.~Paramos,
  Phys.\ Rev.\  D {\bf 72}, 123512 (2005)
  [arXiv:astro-ph/0509547].
\bibitem{Gorini:2008zj}
  V.~Gorini, U.~Moschella, A.~Y.~Kamenshchik, V.~Pasquier and A.~A.~Starobinsky,
  Phys.\ Rev.\  D {\bf 78}, 064064 (2008)
  [arXiv:0807.2740 [astro-ph]].
\bibitem{Gorini:2009em}
  V.~Gorini, A.~Y.~Kamenshchik, U.~Moschella, O.~F.~Piattella and A.~A.~Starobinsky,
  Phys.\ Rev.\  D {\bf 80}, 104038 (2009)
  [arXiv:0909.0866 [gr-qc]].

\bibitem{Shu:1977uc}
  F.~H.~Shu,
  Astrophys.\ J.\  {\bf 214}, 488 (1977).
\bibitem{Bouquet:1985}
S.~Bouquet, M.~R.~Feix, E.~Fijalkow and A.~Munier,
  Astrophys.\ J.\  {\bf 293}, 494 (1985).
\bibitem{Suto:1988}
Y.~Suto and J.~Silk,
  Astrophys.\ J.\  {\bf 326}, 527 (1988).
\bibitem{Lou:2006jn}
  Y.~Q.~Lou and W.~G.~Wang,
  Mon.\ Not.\ Roy.\ Astron.\ Soc.\  {\bf 372}, 885 (2006)
  [arXiv:astro-ph/0608043].
\bibitem{Wang:2008rs}
  W.~G.~Wang and Y.~Q.~Lou,
  Astrophys.\ Space Sci.\  {\bf 315}, 135 (2008)
  [arXiv:0804.2889 [astro-ph]].
\bibitem{Persic:1995ru}
  M.~Persic, P.~Salucci and F.~Stel,
  Mon.\ Not.\ Roy.\ Astron.\ Soc.\  {\bf 281}, 27 (1996)
  [arXiv:astro-ph/9506004].

\end{thebibliography}
\end{document}